\documentclass[journal,twoside,web]{ieeecolor}
\usepackage{tmi}
\usepackage{cite}
\usepackage{amsmath,amssymb,amsfonts}
\usepackage{algorithmic}
\usepackage{graphicx}
\usepackage{textcomp}
\usepackage{url}

\def\BibTeX{{\rm B\kern-.05em{\sc i\kern-.025em b}\kern-.08em
    T\kern-.1667em\lower.7ex\hbox{E}\kern-.125emX}}
\begin{document}

\title{Convolutional Neural Network Model Observers Discount Signal-like Anatomical Structures During Search in Virtual Digital Breast Tomosynthesis Phantoms}

\author{Aditya Jonnalagadda, Bruno B. Barufaldi, Andrew D.A. Maidment, Susan P. Weinstein, Craig K. Abbey, and Miguel P. Eckstein
\thanks{This work has been submitted to the IEEE for possible publication. Copyright may be transferred without notice, after which this version may no longer be accessible.}
\thanks{
This work was supported by the National Institute of Health (R01EB026427), U.S. Army Research Office (W911NF-19-D-0001), American Association of Physicists in Medicine (2020 AAPM Research Seed Funding Grant), Burroughs Wellcome Fund (IRSA 1016451), Terri Brodeur Breast Cancer Foundation (2023 Fellowship) and Susan G. Komen Foundation (CCR231010477). The views and conclusions contained in this document are those of the authors. They should not be interpreted as representing the official policies, either expressed or implied, of the U.S. Government. The U.S. Government is authorized to reproduce and distribute reprints for government purposes, notwithstanding any copyright notation herein.
}
\thanks{Aditya Jonnalagadda is with Department of Electrical and Computer Engineering, University of California, Santa Barbara, Santa Barbara, CA 93106, USA (e-mail: aditya\_jonnalagadda@ucsb.edu).}
\thanks{Bruno B. Barufaldi, Andrew D.A. Maidment and Susan P. Weinstein are with Department of Radiology, University of Pennsylvania, 3400 Spruce Street, Philadelphia, PA 19104, USA.} 
\thanks{Craig K. Abbey and Miguel P. Eckstein are with Department of Psychological and Brain Sciences, University of California, Santa Barbara, Santa Barbara, CA 93106, USA (e-mail: ckabbey@ucsb.edu and migueleckstein@ucsb.edu).}
}



\maketitle

\begin{abstract}

Model observers are computational tools to evaluate and optimize task-based medical image quality. Linear model observers, such as the Channelized Hotelling Observer (CHO), predict human accuracy in detection tasks with a few possible signal locations in clinical phantoms or real anatomic backgrounds. In recent years, Convolutional Neural Networks (CNNs) have been proposed as a new type of model observer. What is not well understood is what CNNs add over the more common linear model observer approaches. We compare the CHO and CNN detection accuracy to the radiologist's accuracy in searching for two types of signals (mass and microcalcification) embedded in 2D/3D breast tomosynthesis phantoms (DBT). We show that the CHO model's accuracy is comparable to the CNN's performance for a location-known-exactly detection task. However, for the search task with 2D/3D DBT phantoms, the CHO's detection accuracy was significantly lower than the CNN accuracy. A comparison to the radiologist's accuracy showed that the CNN but not the CHO could match or exceed the radiologist's accuracy in the 2D microcalcification and 3D mass search conditions. An analysis of the eye position showed that radiologists fixated more often and longer at the locations corresponding to CNN false positives. Most CHO false positives were the phantom's normal anatomy and were not fixated by radiologists. In conclusion, we show that CNNs can be used as an anthropomorphic model observer for the search task for which traditional linear model observers fail due to their inability to discount false positives arising from the anatomical backgrounds. 


\end{abstract}


\section{Introduction}
\label{sec:introduction}

In the field of medical imaging, linear model observers have been extensively used for image quality assessment with signals present at one or a few specified locations (location known exactly, LKE) in filtered Gaussian noise~\cite{Burgess1981EfficiencyOH, Burgess1994StatisticallyDB, p11, Burgess2001HumanOD, Myers1985EffectON, Myers1987AdditionOA, Abbey2001HumanAM, p9}, more complex non-Gaussian backgrounds mimicking anatomical backgrounds~\cite{Rolland1992EffectOR, Castella2008MammographicTS}, and clinical backgrounds~\cite{Burgess2001HumanOD, Eckstein2003AutomatedCE, Castella2008MammographicTS}. When the signal and background statistics are completely known, an ideal observer (IO) implementation is feasible. It provides the upper limit on performance for any observer on a perceptual task and  used to benchmark human performance~\cite{p5,p6,p7,p8,p9}. When the exact statistics are unknown, the IO is not feasible, but approximations are possible~\cite{Kopp2018CNNAM, Gong2020DeeplearningbasedMO}. 
Sub-optimal observers such as Channelized Hotelling Observer (CHO)~\cite{p10} involve a feature extraction stage through a set of linear channels. When the early visual processing in the human visual system is approximated using these linear channels, CHO becomes a better anthropomorphic model observer than an ideal observer~\cite{p11}. Linear model observers have been extended to detection of signals known statistically~\cite{Castella2009MassDO, Lago2020FoveatedMO}, 3D image stacks~\cite{Platisa2011ChannelizedHO} and have become an important part of in-silico virtual clinical trials~\cite{Badano2018EvaluationOD, Samei2020VirtualIT, Abadi2020VirtualCT}.

In recent years, Convolutional Neural Networks (CNN) have also been used as model observers in medical imaging~\cite{p12, p13, r3, Kim2020ACN, Kopp2018CNNAM, Massanes2017EvaluationOC, Lorente2021UnderstandingCB}.
For simple signal-known-exactly and background-known-exactly tasks with Gaussian noise, CNNs provided an excellent approximation to the ideal observer~\cite{p14,p15}.
For the Signal-Known-Exactly (SKE) and Background-Known-Statistically (BKS) tasks, new training strategies for CNN-based anthropomorphic model observers have resulted in good agreement with human observer performance~\cite{Kim2020ACNCT}.
Furthermore, for the SKE-BKS condition, CNNs have been trained with adversarial robust training to generate more human-interpretable features~\cite{r2}.
For a defect forced-localization task, an exploration across different backgrounds, hyperparameters, and loss functions, CNN-based model observer optimized using mean-squared error provided strong agreement with human performance~\cite{r3}.

Even with the larger number of studies that have used CNNs as model observers~\cite{p14,p15,Kim2020ACN,r2,r3}, the advantages of CNNs over conventional model observers are not well-understood. It is also not clear the scenarios for which CNNs should be used instead of linear model observers.   

Our overall goal is to understand the relationship between linear model observers and CNNs across tasks (LKE and search tasks).
In which tasks do linear observers perform at similar accuracy to the CNNs?  In which do linear models fall short of CNN and human detection performance? And, why?

We compare the accuracy of a CNN (segmentation-based nnU-Net~\cite{r7}), two linear model observers (Channelized Hotelling and Filtered Channel Observer) in LKE, and search tasks of one of two signals in Digital Breast Tomosynthesis (DBT) virtual phantoms ~\cite{Pokrajac2012OptimizedGO, Bakic2014RealisticSO}.  The two signals were a small microcalcification-like signal and a larger mass-like signal. 
Furthermore, we compare the model search performances to those attained by twelve radiologists in a search task with the same Digital Breast Tomosynthesis (DBT) data set.  

First, we explore the relationship among the accuracies of the Channelized Hotelling Observer (CHO), Filtered Channel Observer (FCO), and CNN for the commonly used LKE task with increasing number of locations. 

Second, in a more clinically realistic search task, we compare the performance of CHO, FCO,  CNN, and radiologists for a single slice (2D search) and a stack of slices (3D search).  The 2D search in our studies uses the central slice of the 3D signal. We use the 2D search to understand how humans and models search in 2D/3D while controlling for the visual information in the images across conditions.  The 2D vs. 3D search should not be interpreted as comparing real clinical 2D imaging modalities (i.e., mammography or a synthesized 2D image from the 3D image stack) vs. 3D modalities (Digital Breast Tomosynthesis). 

Third, we use the radiologists' eye-fixation data recorded during the visual search to relate fixation locations and times to the signal probability/response maps of the CNN and linear model observers. The goal is to understand whether radiologists and models select similar locations as potential signals and shed some light on the differences between the CNN and linear model observers.


%

\section{Materials and Study}

\subsection{Dataset}
We used DBT phantoms as test images for the radiologist study and evaluated CNN and linear model observers. The phantoms were generated by the OpenVCT virtual breast imaging tool from the University of Pennsylvania~\cite{Pokrajac2012OptimizedGO, Bakic2014RealisticSO}. 
This tool generates full phantom DBT images, including different tissues (skin, Cooper's ligaments, adipose, and
glandular) in a realistic manner.
Breast phantoms are generated using a $100$ um voxel size ($78.4\times205.3\times63.3$ $mm^{3}$). Tomosynthesis projections are simulated following the acquisition geometry of a clinical system (Selenia Dimensions, Hologic Inc., Bedford, MA). Reconstructions are generated using a commercially-available library (Briona, Real-Time Tomography, Villanova, PA). In total, the image dataset of each phantom comprises $64$ slices of size $2048\times1792$ pixels.
The dataset contains two types of signals: 1. a small lesion similar to microcalcification and 2. a large lesion similar to a mass.
We have a dataset of about 500 phantoms with microcalcification, 500 phantoms with mass, and 500 signal-absent phantoms.
For 2D experiments, we used a single slice of the 3D phantom, which contained the signal's center as the signal-present phantom. The signal-absent 2D images were also a slice from signal-absent 3D phantoms.

\subsection{Radiologist study}

\begin{figure}[!t]
    \centerline{\includegraphics[width=\columnwidth]{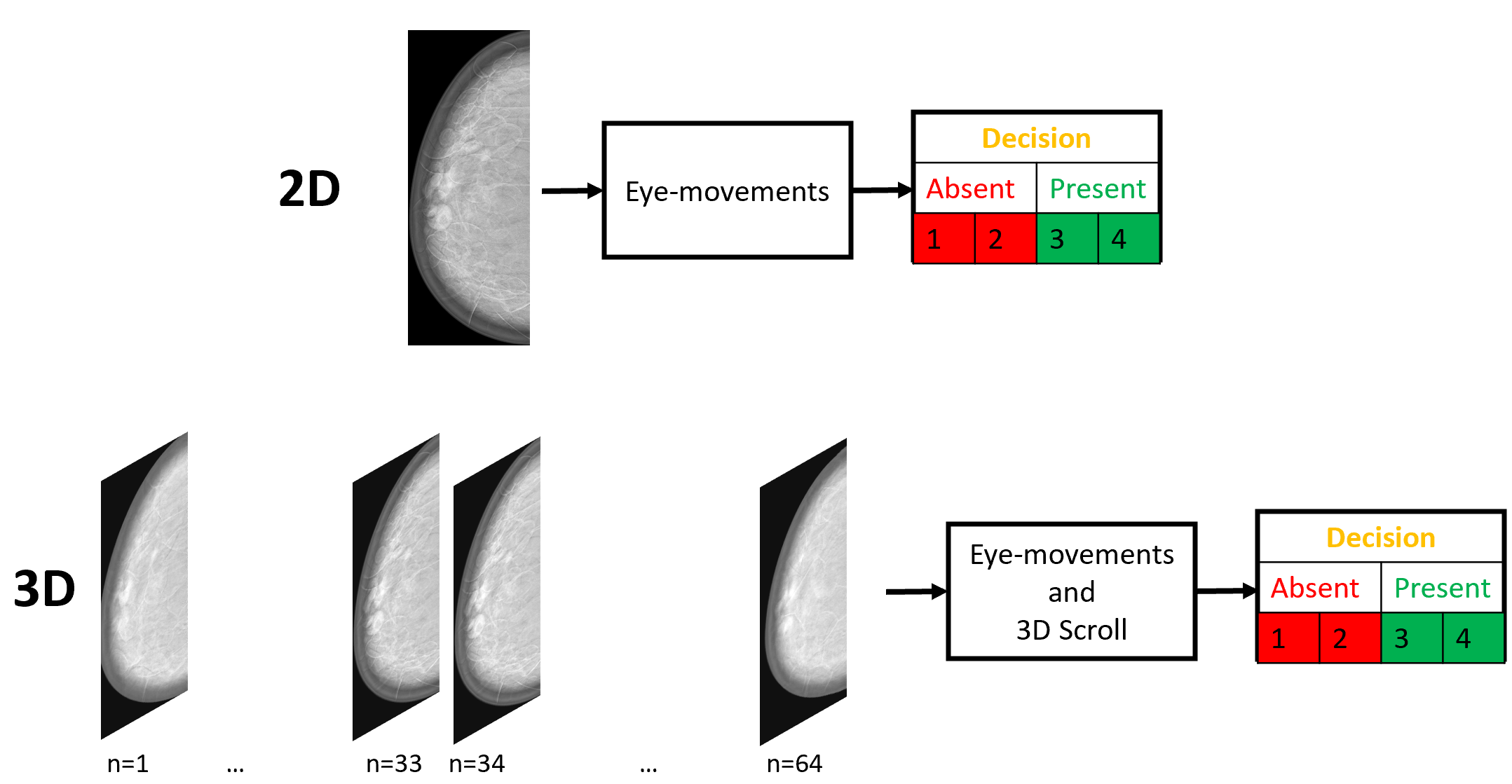}}
    \caption{\textbf{Radiologist study:} Twelve radiologists participated in the study. Each was shown a total of 28 2D phantoms and 28 3D phantoms. Half of each set had a signal present and another half with signal absent. Half of the signal-present images contained a microcalcification signal, while the other half had a mass signal. After making a decision, radiologists used a 4-point decision confidence scale where 4 corresponds to strong confidence of signal presence, 3 corresponds to moderate confidence of signal presence, 2 corresponds to moderate confidence of signal absence, and 1 corresponds to strong confidence of signal absence. For the 2D DBT, only eye movements were made, and for 3D, eye movements and scrolling across slices were possible.}
    \label{fig0}
\end{figure}

Twelve radiologists participated in the study. They sat $75\text{ - }95$ cms away from a vertical medical-grade monitor in a darkened room.
Stimuli were displayed on a 5Mpx grayscale DICOM calibrated monitor ($2560\times2048$ pixels), keeping their aspect ratio.
In the study, each radiologist saw $28$ signal-present (microcalcification or mass) and $28$ signal-absent trials, out of which half were 2D trials and the other half were 3D trials.
The prevalence between microcalcifications and masses was $50\%$. 
Microcalcification was simulated as a solid sphere with a diameter of 0.3 mm, and mass was simulated as a combination of 3D ellipsoids with an average diameter of 7 mm.
Microcalcifications and masses subtended visual angles of 0.06 and 0.5 degrees, respectively.
All four conditions were randomly
intermixed. An eye tracker recorded the participant’s eye movements in real-time at a frequency of $500$ Hz (EyeLink Portable Duo, SR Research). We also recorded the timing and slices of the scrolling for the 3D search. We used Psychtoolbox to develop this experiment~\cite{psychtoolbox07}.
Each radiologist only saw a subset of the phantoms. Area-under-the-curve (AUC) is computed for each radiologist.

%

\section{Model observers}


We trained three model observers: 1. Channelized Hotelling Observer (CHO) with Gabor Channels~\cite{Eckstein2000APG, p2, Zhang2004AutomatedOO, DIDSR_MO_implementation}, 2. Filtered Channel Observer (FCO)~\cite{Daz2015DerivationOA, DIDSR_MO_implementation} and 3. Convolution Neural Network (CNN)~\cite{r7, r8}.
We used a set of approximately $1500$ phantoms for training the model observers.

\subsection{Gabor channel computation}
We used Gabor channels with $N_{o}$ orientations, $N_{p}$ phases, and $N_{f}$ spatial frequencies, resulting in a set of $N_{o}\times N_{p}\times N_{f}$ Gabor channels, represented by $C^{Gabor}$.
\begin{equation}
    i^{'} = i\cos{(\theta_{k})}+j\sin{(\theta_{k})}
\end{equation}
where $i$, $j$ are the x-y coordinates, $\theta_{k}$ is the $k$th orientation and $i^{'}$ is the transformed x coordinate. And $k$th channel of the Gabor channel array, $C_{k}^{Gabor}$, is computed using,
\begin{equation}
    C_{k}^{Gabor}(i,j) = Gaus(i,j)*\cos{(2 \pi*fc_{l}*i^{'}   +\beta_{p})}
    \label{G3}
\end{equation}
where $Gaus(i,j)$ is the Gaussian envelope, $fc_{l}$ is the $l$th central frequency and $\beta_{p}$ is the $p$th phase.
We created Gabor channels with $8$ orientations, $2$ phases, and $5$ spatial frequencies.

\subsection{Channelized Hotelling observer (CHO)}

The Chanellized Hotelling computes the d' maximizing weight of channels, which is optimal for Gaussian noise processes but not for backgrounds with higher-order moments. Using dot product, we compute the response of the training image crops to the Gabor channels (from Equation~\ref{G3}),
\begin{equation}
    V = (C^{Gabor})^{T}S
    \label{CHO_1}
\end{equation}
where $S_{P}$ and $S_{A}$ are the arrays of signal-present and signal-absent training crops, respectively, and $S \in \{S_{P}, S_{A}\}$; 
$V_{P}$ and $V_{A}$ are the arrays of signal-present and signal-absent responses respectively and $V \in \{V_{P}, V_{A}\}$;
$T$ is the transpose operation.

The signal-present and signal-absent covariance matrices, $K_{P}$ and $K_{A}$, are computed using the response arrays $V_{P}$ and $V_{A}$.
\begin{equation}
    S_{ch} = (1/N_{P})*\sum_{N_{P}}V_{P} - (1/N_{A})*\sum_{N_{A}}V_{A}
\end{equation}
\begin{equation}
    K = (K_{P}+K_{A})/2
    \label{CHO_2}
\end{equation}
Weights of the linear CHO ($W_{CHO}$) are computed using a dot product, 
\begin{equation}
    W_{CHO} = S_{ch}^{T}K^{-1}
    \label{CHO_3}
\end{equation}
Response of test images ($\textit{g}$) to the CHO is computed using convolution in the frequency domain. 
\begin{equation}
    \lambda = W_{CHO}^{T}.g_{ch}
    \label{CHO_4}
\end{equation}
Where $g_{ch}$ refers to the response of the test images to the Gabor channels, $C^{Gabor}$.

\subsection{Filtered channel observer}
The Filtered Channel Observer (FCO)~\cite{Daz2015DerivationOA} is a linear model observer based on convolution channels capable of modeling irregularly-shaped signals with fewer directional channels.
Constructing the weights of FCO follows a similar procedure to that of CHO.
The implementation differs only in the construction of the channel templates.

Signal mean ($signal_{mean}$) is the difference between the mean signal-present image and the mean signal-absent image. Furthermore, using the Fast Fourier Transform (FFT), the signal mean ($signal_{mean}$) is transformed into the frequency domain ($signal_{FFT}$).
\begin{equation}
    signal_{mean} = (1/N_{P})*\sum_{N_{P}}S_{P} - (1/N_{A})*\sum_{N_{A}}S_{A}
\end{equation}
\begin{equation}
    signal_{FFT} = \text{FFT}(signal_{mean})
\end{equation}
where $S_{P}$ and $S_{A}$ are signal-present and signal-absent training crops respectively, and $N_{P}$ and $N_{A}$ represent the number of signal-present and signal-absent training images.

Gabor channel array (Equation~\ref{G3}) is transformed using the signal mean in the frequency domain and is then converted back into the spacial domain using inverse Fast Fourier Transform ($\text{FFT}^{-1}$), followed by \textit{L-2} normalization.
\begin{equation}
    C_{k}^{'} = (1/nxy)*\text{ABS}(\text{FFT}(C_{k}^{Gabor}))^{2}
\end{equation}

\begin{equation}
    C_{k}^{FCO} = \text{FFT}^{-1}(  C_{k}^{'}  *    signal_{FFT}  )
\end{equation}

\begin{equation}
    C_{k}^{FCO} = C_{k}^{FCO} / sqrt(\sum_{x}\sum_{y}(C_{k}^{FCO})^{2})
\end{equation}

where $C_{k}^{'}$ is the power spectral density of the $k^{th}$ channel $C_{k}^{Gabor}$, $nxy$ is the spatial size of the channel $C_{k}^{Gabor}$, $C_{k}^{FCO}$ is the $k^{th}$ channel of the FCO model.

After the computation of the FCO channels, the template is computed similarly to the CHO template by replacing $C^{Gabor}$ with $C^{FCO}$ in Equation~\ref{CHO_1} followed by computation of template weights and the response map using Equations~\ref{CHO_2},~\ref{CHO_3} and~\ref{CHO_4}.

\begin{figure}[!t]
\centerline{\includegraphics[width=0.95\columnwidth]{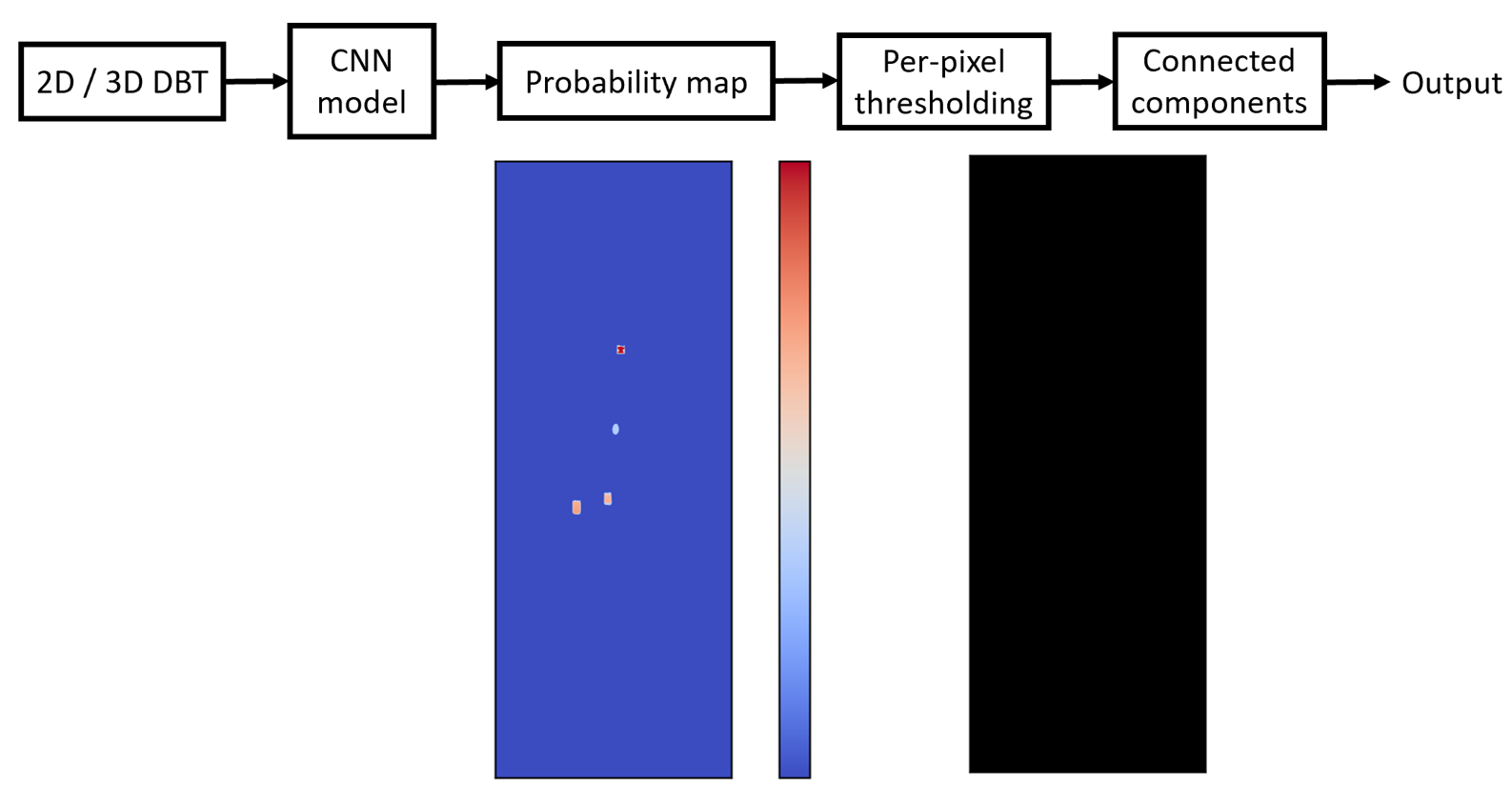}}
\caption{\textbf{CNN model:} Four different CNN models are trained for two modalities (2D/3D) and two signal types (microcalcification/mass). During the test phase, the segmentation-based CNN produces a probability map with a probability value assigned to each pixel, representing its probability of being the signal pixel. A per-pixel threshold computed using the validation set is applied to the CNN output to convert the probability map into a binary map. Connected components are computed using 8 and 26 connectivity for 2D and 3D, respectively, on the resultant binary map.}
\label{fig10_CNN}
\end{figure}

\begin{figure}[!t]
\centerline{\includegraphics[width=0.95\columnwidth]{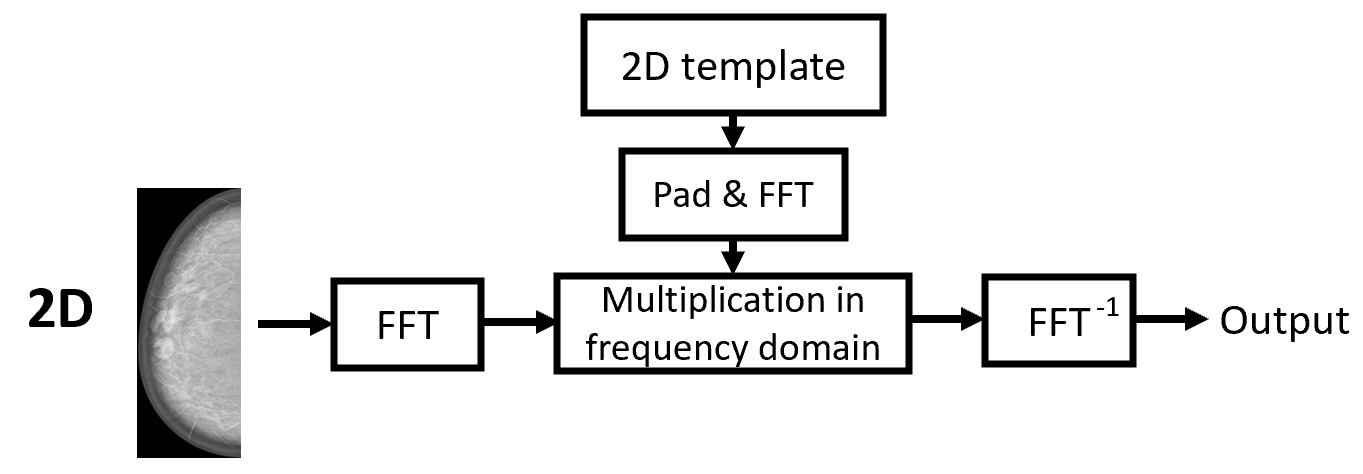}}
\caption{\textbf{2D search with CHO and FCO:} During the testing phase, the 2D template is padded to the size of the 2D input. After taking the Fast Fourier Transform (FFT) of both the input and template, the convolution of the two is performed in the Fourier domain by performing their multiplication. The final response map is generated by taking the inverse FFT of the convolution output in the frequency domain.}
\label{fig10_CHO}
\end{figure}

\subsection{Convolution neural network}
U-Net~\cite{r8} 
is a CNN architecture used for performing image segmentation. 
It consists of two stages: encoder and decoder. The encoder stage gradually reduces the spatial dimensions while increasing the feature dimensionality. The decoder stage performs the opposite operation, gradually increasing the spatial dimension while decreasing the feature dimensionality, resulting in an output map of the same spatial size as the input image.
Thus, the model can be trained to produce a segmentation map as the final output map.
NN-UNet~\cite{r7} is a U-Net~\cite{r8} based convolution neural network with automatic hyper-parameter computations.
It was originally designed for the Medical Segmentation Decathlon~\cite{a7} challenge, consisting of segmentation tasks corresponding to different target regions in the human body, and it was evaluated on $23$ datasets in the biomedical domain using a fully automated pipeline.
We choose to use this architecture due to its optimized hyper-parameter computation and successful applicability to a wide variety of tasks.
It is a segmentation-based CNN that outputs a probability map signifying the likelihood of each pixel being a lesion or background.
A segmentation-based CNN architecture provides a basic output visualization without the necessity of additional interpretability techniques~\cite{r4,r5,r6}.

One limitation noted is that we only used mean signal information and the center of lesion coordinates mapped to reconstructions to indicate the signal-present phantoms; the mask of the ground-truth lesion should be used to denote the pixels belonging to the lesion. However, using the center location and the bounding boxes of lesions (i.e., lesion dimensions) provides sufficient information to inform the actual location of the lesions.
Although a custom CNN designed for this task might have resulted in better performance, it might not be readily generalizable/applicable for phantoms designed slightly differently. For this reason, we expect the results to be more generalizable by using a standard architecture that works for different types of medical images.

In the encoder, downsampling is performed using strided convolutions; in the decoder, upsampling is performed using transposed convolutions. 
Each resolution stage in the encoder/decoder consists of two computational blocks, and each computational block consists of convolution followed by instance normalization~\cite{Ulyanov2016InstanceNT} and leaky-ReLU~\cite{Maas2013RectifierNI}.
For the 2D modality, a 2D architecture is used where the number of channels in the input image is just one slice.
For the 3D modality, there are two options: 
1. 3D full-resolution U-Net, which operates on full-resolution input, 
2. 3D U-Net cascade, where first coarse segmentation is performed on low-resolution data, then refined by a 3D U-Net operating on full-resolution data.
We use the option of 3D full-resolution U-Net.

After separating the training dataset into five folds, five CNN models are trained by choosing four folds for training and the fifth for validation.
We train each of the five models for $1000$ epochs.
Stochastic Gradient Descent optimizer with a learning rate of $0.01$ and nestrov momentum~\cite{Nesterov1983AMF, Sutskever2013OnTI} of $0.09$ are used.
The learning rate is controlled using a 'polyLR' learning schedule, where a polynomial function is used to decay the learning rate of each parameter group.
After training, the five models are combined to form the ensemble model, which is then used for testing.
All four CNN models (2D microcalcification, 2D mass, 3D microcalcification, and 3D mass) are trained using a similar procedure.


\section{Implementation details}

\subsection{Gabor channels}






We created Gabor channels with $8$ orientations, $2$ phases, and $5$ spatial frequencies.
The eight orientations are equally spaced apart between $0$ to $\pi$ i.e., $N_{o} = 0, \pi/8, ..., 7\pi/8$.
The two phases are orthogonal, i.e., $N_{p} = 0, \pi/2$.
Five spatial frequencies are used for channel generation, i.e., $N_{f} = 4, 8, 16, 32, 64$ pixels per cycle. 
For a monitor distance of $85$ cms, it is equivalent to a spatial frequency in cycles per degree of $16, 8, 4, 2, 1$, respectively.
Image crops used for training are of size $101 \text{ pixels} \times 101 \text{ pixels}$.
Using these orientations, phases, and spatial frequencies, a Gabor channel array of size $101 \times 101 \times 80$ with $80$ channels is created.

\subsection{Linear model observers CHO and FCO}

\subsubsection{Training the 2D template}
Both 2D and 3D templates are of size $101$ pixels in width and $101$ pixels in height.
In terms of depth, the 2D template consists of a single slice, and the 3D template consists of $17$ slices, i.e., eight slices on either side of the central slice.
The training dataset consists of $576$ DBT phantoms, each of size $2048\times1792\times64$, for each microcalcification, mass, and signal-absent category.
For the 2D case, for each signal-present phantom, a crop of size $101\times101$ is obtained from the signal location on the central slice, resulting in an array of size $101\times101\times576$.
For the signal-absent images, ten random crops are obtained from each phantom, resulting in an array of size $101\times101\times5760$.
The eighty-channel Gabor array is of size $101\times101\times80$.
Channel response is computed using the Gabor array for both signal-present and signal-absent arrays, resulting in response arrays of sizes $80\times576$ and $80\times5760$, respectively.
The mean signal of size $80\times1$  is obtained by computing the means of both these arrays and taking their difference.
We compute the covariance matrices for signal-present and signal-absent separately, each of which is of size $80\times80$, and then obtain the mean covariance matrix by taking the mean of both signal-present and signal-absent covariance matrices.
Finally, the template weights are computed according to equation 3 using the mean signal and the pseudo inverse of the mean covariance matrix.

\subsubsection{Training of 3D template with optimized slice weights}
For training the 3D templates, we train the individual 2D templates for all the $17$ slices as per the procedure described above.
It would be sub-optimal to give uniform weights to all the templates and combine them, so we train the weights according to how the template responses need to be combined. 

We follow a similar procedure to how we trained the 2D template except that here the different 2D templates are similar to the channels present during the generation of the 2D template, i.e., the mean signal is of size $101\times101\times17$. The mean covariance matrix is of size $17\times17$.
As a result, for each of the 3D templates, we have $17$ 2D templates, each of size $101\times101$ and $17$ scalar weights, to combine the 2D templates optimally. 

\subsubsection{Testing}
For both CHO and FCO, Figure~\ref{fig10_CHO} illustrates the testing procedure for 2D and 3D inputs.
We compute the convolution between the template and the input in the frequency domain by computing the Fast Fourier Transform (FFT) of the input phantom and the padded template and multiplying them in the frequency domain.
We obtain the final response by performing the inverse FFT of the signal's and template's product in the frequency domain.
False positives in the skin tissue and the edge of the phantom degrade the search performance of CHO and FCO. To boost their performance, we mask these locations by multiplying the response map with a binary map.
We use the largest value in the resultant map to represent the phantom.

\subsection{Convolution neural network}
\subsubsection{Pre-processing}
To aid the training of the segmentation-based CNN, we constructed binary masks representing the signal locations using foreground pixels and the noise locations with background pixels.
The dataset only provides the coordinates of the signal location but not the binary masks needed for the training.
Using the template of the mean signal, we constructed these masks. 
Constructing more accurate masks using costly manual human annotations might be possible.
Approximate binary masks result in more challenging CNN training with the trade-off of cheap binary mask generation.

\subsubsection{Training}
We trained four CNN models for the two signal types (microcalcification and mass) and the two modalities (2D and 3D).
For the 2D case, $1500$ single slice samples of size $2048\times793\times1$ are used for training, where the training samples are generated using the central slice ($C_{slice}$), slice before the central slice  ($C_{slice}-1$) and slice after the central slice  ($C_{slice}+1$) of the $500$ 3D signal-present training phantoms.
For the 3D case, $500$ samples of size $380\times380\times64$ are used for training, where the $380\times380$ crops are generated from the full spatial size of $2048\times793$ and a single sample is obtained from each of the $500$ training phantoms. The cropping is done to save training time. Uncropped full phantoms were presented as input to the model during testing. This change works because it is a segmentation-based CNN.
Each CNN model is an ensemble of five individually trained CNN models, where each individual CNN is trained by splitting the training dataset into five folds. Each of the individual CNNs uses four folds for training and the remaining for validation.
We train each of the individual CNN models for $1000$ epochs.
Each epoch took approximately $240$ seconds and $800$ seconds for the 2D and 3D, respectively, on a single $12$ GB Nvidia GPU.
Therefore, training the 2D models approximately takes $240\times1000\times5\times2$ seconds, which translates to about $28$ days on a single GPU, accounting for $1000$ epochs for each model, $5$ folds for each model and $2$ models one for microcalcification and mass.
Training the 3D models takes approximately $93$ days on a single GPU. 
We used four GPUs to decrease the total training time.
We did not change parameters like the number of epochs to reduce training time to maintain the generalization ability.

\subsubsection{Post-processing}
The presence or absence of the signal cannot be directly determined using the probability map output of the CNN.
We use a two-step process of per-pixel thresholding and connected components to make it interpretable.
Under per-pixel thresholding, a threshold optimized using the validation set is used to classify each pixel as a foreground (signal) pixel or a background (noise) pixel by comparing the probability value at that pixel against the pre-computed threshold.
We compute the connected components using 8 and 26 connectivity, using the understanding that the presence of the signal will lead to a cluster of foreground pixels.
We compute the volume of the largest connected component and use it as the response of the entire phantom. If the actual signal is present, this value will be higher.

\subsubsection{Testing}
During the testing phase, the input is passed through the ensemble CNN, followed by per-pixel thresholding and the computation of the connected components.
The per-pixel threshold is computed using a validation set of $50$ signal-present and $50$ signal-absent phantoms, and it is individually computed for each of the four models, two signal types, and two modalities.
Since the CNN output is a probability map, the value of each output pixel varies between $0$ and $1$. In an ideal scenario, each foreground (signal) pixel will have a value of $1$, and each background (noise) pixel will have a value of $0$. 
By reducing the per-pixel threshold in steps of $0.05$, the threshold at which the AUC for the validation set is maximized is used as the per-pixel threshold for the test set. 
By using 8-connectivity for 2D and 26-connectivity for 3D, we compute all the connected components in the CNN output map. The largest connected component is assumed to represent the potential lesion location, and the number of pixels in the largest connected component is used as the network confidence on the presence of the lesion and is used as the representative of this input in the computation of AUC. No further thresholding of the connected component volume is needed, as Area-under-the-curve (AUC) is used as the figure-of-merit.

\subsection{Bootstrap statistics}
We used bootstrap statistics~\cite{Efron1979BootstrapMA} to check for the statistical significance of the difference between the observers. During the experimental study, each of the twelve radiologists saw seven signal-absent and seven signal-present phantoms under each condition (microcalcification/mass and 2D/3D). The phantoms seen by the radiologists had partial overlap. The set of phantoms used for the study was sampled from a set of unique phantoms consisting of $27$ signal-absent and $14$ signal-present phantoms. 

\subsubsection{Search task}
Statistics are computed using $20,000$ bootstrap iterations.
At each bootstrap iteration, a phantom set is generated by sampling from the set of unique phantoms with replacement, where signal-present and signal-absent are sampled separately.
The set of twelve radiologists is sampled with replacement to generate a reader set.
For each radiologist in the reader set, phantoms from the phantom set are assigned if they were seen by that radiologist during the study. This also ensures that a rating is available for all the assigned phantoms.
As a result, if any of the radiologists in the radiologist set are assigned less than six signal-present or six signal-absent phantoms, the bootstrap iteration is discarded and is not counted towards the $20,000$ iterations.
If the bootstrap iteration is valid, parametric AUCs are computed.
This sampled phantom set is then used to compute the parametric AUC of the model observers (CNN, CHO, and FCO).  
For computing the radiologist AUC, the parametric AUC for each radiologist in the reader set is averaged to get the overall radiologist AUC.
The difference in AUCs between different observers is stored at the end of each iteration for computing the final bootstrap statistics.
At the end of the $20,000$ iterations, for each of the distribution of observer AUC differences, the percentile value of the zero difference in the distribution of bootstrap differences is computed to check for the statistical significance of the difference between the observers involved in the computation of differences.

\subsubsection{Time-spent Analysis}
A similar procedure is used for computing the bootstrap statistics between the model observers, where CNN, CHO, and FCO are compared against each other.
Since the analysis involves only the signal-absent phantoms, the phantom set is generated from the unique set of signal-absent phantoms only.
At each of the bootstrap iterations, after generating the phantom set and the reader set, the average percentage of time spent is computed for each of the model observers, and the difference between them is stored.
At the end of the $20,000$ iterations, similar to the search task, the percentile value of the zero difference in the distribution of bootstrap differences is computed to check for the statistical significance of the difference between the model observers involved in computing the difference.

\section{Results}

We compared the detection accuracy for both signals from all three model observers for the LKE and search tasks.  
The location-known-exactly (LKE) task consisted of comparing the potential locations of signal-present phantom against the potential locations in the signal-absent phantom, where the signal location is always part of the potential locations of the signal-present phantom.
Phantom level response of yes/no (signal presence/absence) is obtained based on the maximum response among the potential locations.
LKE task is repeated for different numbers of potential locations per image.
In the search task, the signal is located at a random location in the image.   Linear model observers convolve the image with the signal templates by performing multiplication in the frequency domain, and the yes/no (signal presence/absence) is obtained based on the maximum response location in the convolution output. The search task is more similar to the perceptual task of radiologists in the clinic.  

\subsection{Location Known Exactly (LKE) Task}

\begin{figure}[!t]
\centerline{\includegraphics[width=\columnwidth]{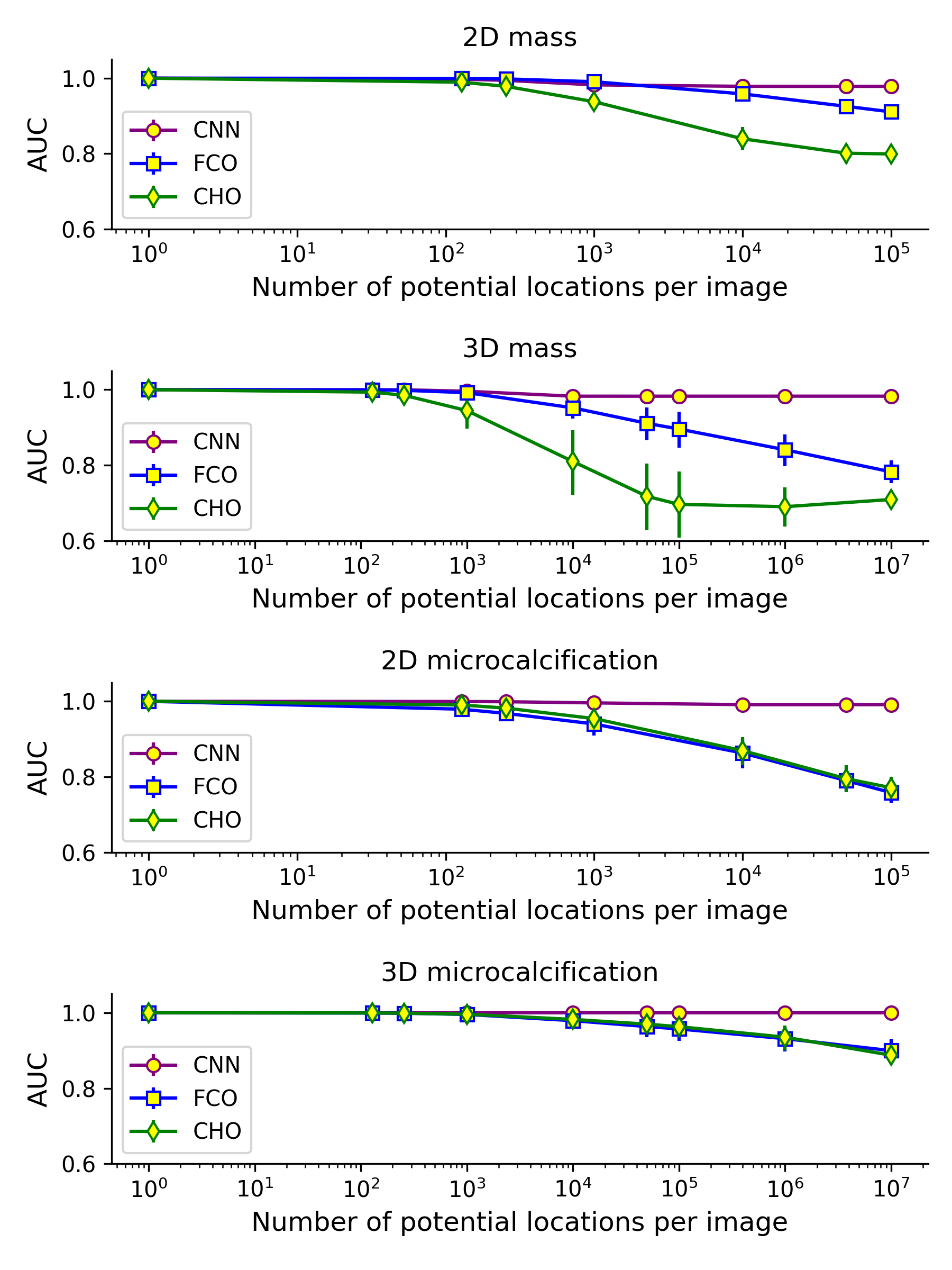}}
\caption{\textbf{Linear model observers vs. CNN in LKE task:} Accuracy (area under the ROC, AUC) for models for the microcalcification (CALC) and mass signals. CNN: Convolutional Neural Network, CHO: Channelized Hotelling Observer; FCO: Filtered Channel Observer}
\label{fig_LKE_Search}
\end{figure}

Each row of Figure~\ref{fig_LKE_Search} shows the performance comparison between the linear model observers (CHO, FCO) and the CNN.
Performance comparisons are shown for 2D mass, 3D mass, 2D microcalcification, and 3D microcalcifications in the first, second, third, and fourth rows, respectively.
LKE performance for each condition is computed for a different number of potential locations per image (N), where N takes the values $1$, $128$, $256$, $1,000$, $10,000$, $50,000$, $100,000$ for the 2D case and N takes the values $1$, $128$, $256$, $1,000$, $10,000$, $50,000$, $100,000$, $1,000,000$, $10,000,000$ for the 3D case.
For both signal types (microcalcification and mass), the performance of the LKE task is at or close to the ceiling ($1.0$) for fewer potential locations per image, and the AUC gradually drops towards the search performance as the number of potential locations per image increases.

For each value of N, the corresponding number of responses are randomly collected from the response map of the template, which is obtained by the model observer operating on the input phantom.
For the signal-present phantom, the value from the signal location is computed as the maximum response in the neighborhood of the signal location. 
Here, the neighborhood of $51 \times 51$ is used for the 2D case and a neighborhood of $51 \times 51 \times 7$ is used for the 3D case.
The N locations on the response map are selected randomly without replacement.
$10,000$ iterations are used to generate the $95\%$ confidence intervals for each data point in Figure~\ref{fig_LKE_Search}.



\textbf{2D case: } First and third rows of Figure~\ref{fig_LKE_Search} show the LKE performance of the 2D mass and 2D microcalcification, respectively. 
As the number of potential locations per image increases, the LKE performance becomes equivalent to the search task.
For both types of signals (microcalcification and mass), the performance of the LKE task is close to the ceiling ($0.999$) at N = $1$ for both the linear model observers (CHO and FCO) and CNN. 
As the value of N increases, the LKE task becomes relatively similar to a search task, and the CNN outperforms both CHO and FCO.
For microcalcification ($\Delta AUC_{\text{$N_{A}$ vs. $N_{B}$}}$ = $0.229$ for CHO and $\Delta AUC_{\text{$N_{A}$ vs. $N_{B}$}}$ = $0.242$ for FCO) and 
mass ($\Delta AUC_{\text{$N_{A}$ vs. $N_{B}$}}$ = $0.2$ for CHO and $\Delta AUC_{\text{$N_{A}$ vs. $N_{B}$}}$ = $0.089$ for FCO), where $N_{A}$ and $N_{B}$ are $1$ and $100,000$, respectively.
This suggests that the linear model observer responses to the normal-anatomy generated many false positives when there are a high number of potential locations per image in LKE task. 



In contrast, the AUC, for CNN, for each type of signal remains approximately the same across different values of the number of potential locations per image of the LKE task, $\Delta AUC_{\text{$N_{A}$ vs. $N_{B}$}}$ = $0.009$ for microcalcification and $\Delta AUC_{\text{$N_{A}$ vs. $N_{B}$}}$ = $0.021$ for mass, where $N_{A}$ and $N_{B}$ are $1$ and $100,000$ respectively. 
Thus, the normal DBT background in the LKE task with more potential locations per image did not generate as many false positives as the linear models. This finding might suggest that CNN learns to discount normal-anatomy structures that are potentially confusable with the signal.

\textbf{3D case: }  The results for 3D parallel those of 2D. 
 The second and fourth rows of Figure~\ref{fig_LKE_Search} show the LKE performance comparison of 3D mass and 3D microcalcification. 
For the CHO and FCO model observers, the accuracy was lower for both signal types as the number of potential locations per image increased (for the microcalcification 
$\Delta AUC_{\text{$N_{A}$ vs. $N_{C}$}} = 0.112$ for the CHO and 
$\Delta AUC_{\text{$N_{A}$ vs. $N_{C}$}} = 0.1$ for the FCO; 
for the mass 
$\Delta AUC_{\text{$N_{A}$ vs. $N_{C}$}} = 0.29$ for the CHO and 
$\Delta AUC_{\text{$N_{A}$ vs. $N_{C}$}} = 0.217$ for the FCO), where $N_{A}$ and $N_{C}$ are $1$ and $10,000,000$, respectively.

AUC of the CNN for each type of signal remains similar across different number of potential locations per image of the LKE task for the microcalcification ($AUC = 0.999$), and a small decrease in accuracy for the mass ($\Delta AUC_{\text{$N_{A}$ vs. $N_{C}$}} = 0.017$), where $N_{A}$ and $N_{C}$ are $1$ and $10,000,000$ respectively.

A critical finding that applies to both 2D and 3D tasks is that the linear model observers resulted in similar accuracy to CNN when there are fewer number of potential locations per image but displayed lower accuracy when there are higher number of potential locations per image.

\subsection{Model observers vs. Radiologists for Search Task}
\begin{figure}[!t]
\centerline{\includegraphics[width=\columnwidth]{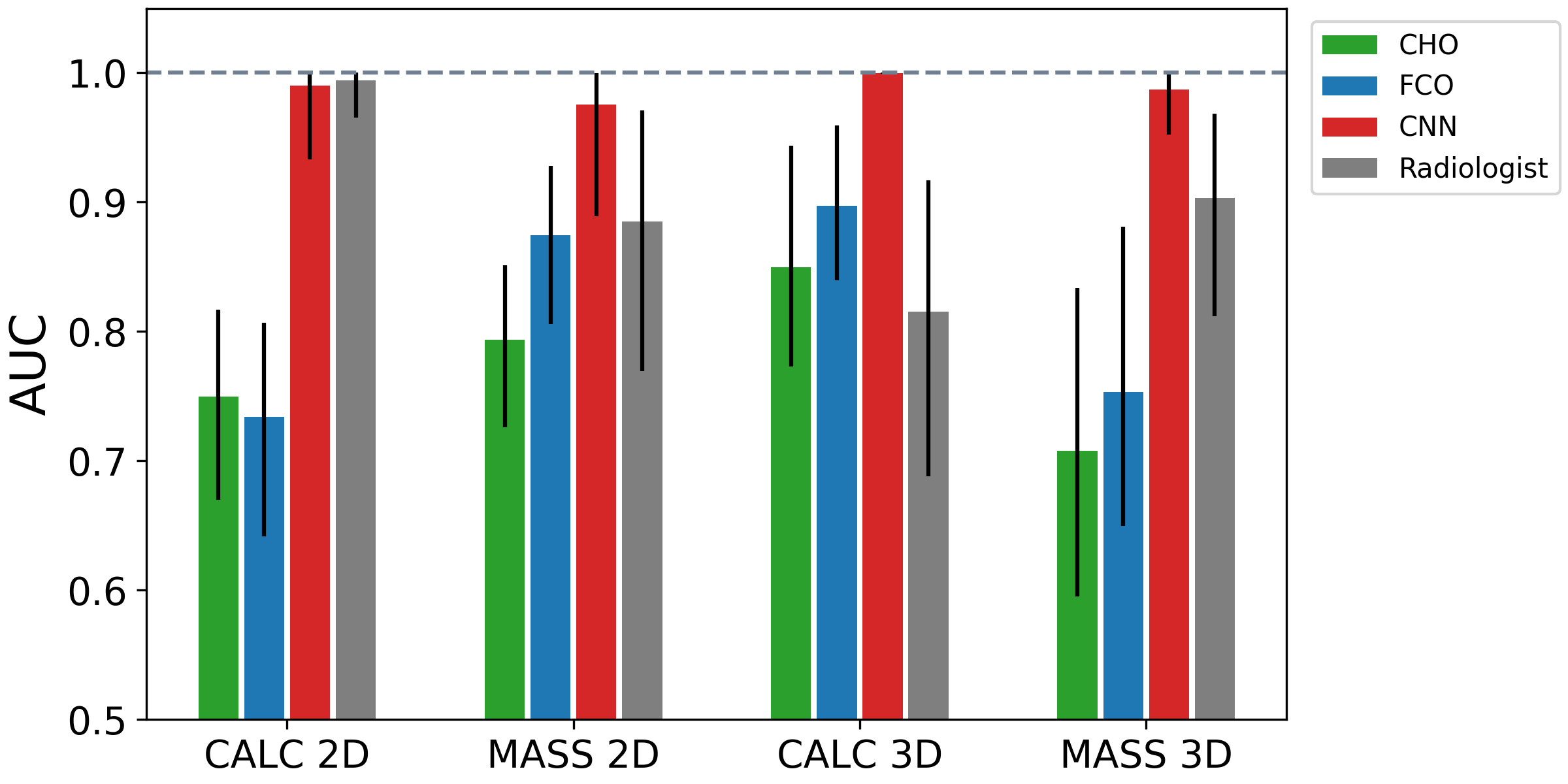}}
\caption{\textbf{CNN, linear model observes (CHO, FCO) and radiologists in 2D and 3D search:} For CALC signal, radiologists underperformed in 3D search, whereas model observers improved their performance from 2D to 3D. For MASS, where the CNN performance still increased from 2D to 3D, CHO and FCO underperformed in 3D, where a better integration across slices is needed.}
\label{fig_2D_3D_Search}
\end{figure}

Next, we compared linear model observers and CNN to radiologists for the search task.  No radiologist data was collected for the LKE task. Figure~\ref{fig_2D_3D_Search} shows accuracy for 2D and 3D search tasks for radiologists and model observers. 

Radiologist AUC was significantly higher ($p < 1e-4$)  than the CHO and FCO for 2D microcalcification and significantly higher ($p < 0.05$) than the CHO for 3D mass. It is not significantly lower or higher for all other conditions.  
In contrast, the CNN attained significantly higher AUC ($p < 0.005$) than the CHO and FCO for all conditions except for the FCO and 2D mass. 
CNN also achieved significantly ($p < 0.025$) better AUC than radiologists for 3D conditions, and it is not significantly lower or higher for 2D conditions. 

\textbf{2D vs. 3D search: }
Performance of the CNN shows a small increase from 2D to 3D, which can be attributed to the larger extent of the 3D signal template relative to the 2D signal template ($\Delta AUC_{\text{2D vs. 3D}} = 0.01$ for microcalcification and $\Delta AUC_{\text{2D vs. 3D}} = 0.012$ for mass).

Accuracy from 2D to 3D for the CHO and FCO models increased for the microcalcification but decreased for the mass.  

Neither the CNN nor the linear model observers capture the reduction of radiologists' accuracy from 2D to 3D for microcalcifications ($p < 0.001$). 
 

\subsection{Models False Positives and Radiologist Fixations}
\label{time_spent_analysis_results_section}
To assess why radiologists might be attaining a higher accuracy than the linear model observers during visual search, we related the models' false positives and the locations radiologists considered suspicious. As a proxy to suspicious locations, we analyzed locations in the signal absent 3D images where radiologists scrutinized and fixated as measured with an infra-red video-based eye tracker. 

Multiple radiologists searched for microcalcification and mass in each of the $27$ signal-absent phantoms. The total number of radiologists varied between $4$-$8$ for each phantom. During their search, the eye tracker recorded the eye movements of the radiologists as they searched the slice and scrolled through the stack of slices. We used a Gaussian filter of size $45\times45\times3$ to smooth the response maps from model observers and time-spent maps from radiologists. Along with the fixation location, we also have access to the information time spent at each fixation location from the eye-tracker. As a result, we can assess whether the locations where the radiologists looked for the longest time are related to the high response values from the model observer response maps. To evaluate the relationship only in the regions where the phantom is present, a binary mask removes all fixations falling outside the actual phantom region. Our analysis focused on the signal-absent phantoms.

This analysis selects the top $1\%$ response score locations from the model observer response map. We then obtained time spent by the radiologists fixating at those locations. To normalize across phantoms, we report the radiologists' percentage of time spent fixating these locations out of the total time spent on this phantom. Thus, the measure combines fixations across radiologists for a given image. 
\begin{figure}[!t]
\centering
\centerline{\includegraphics[width=\columnwidth]
{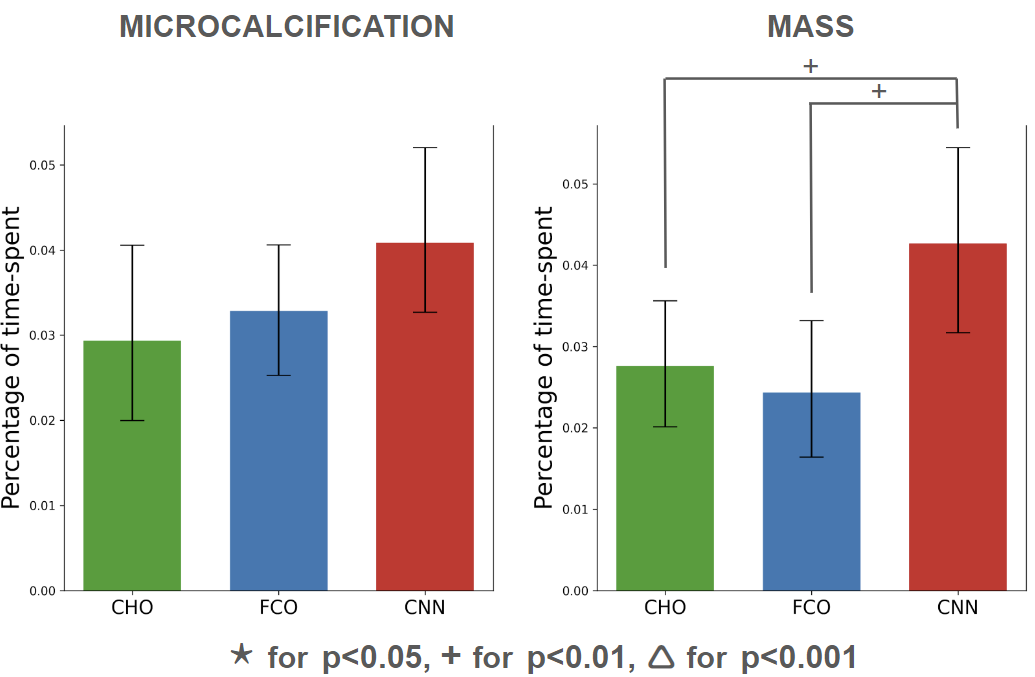}}
\caption{
\textbf{Time-spent by radiologists at top model observer response locations:} 
Percentage of time spent by the radiologists corresponding to top $1\%$ locations of the model observer response map.
}
\label{time_spent_analysis}
\end{figure}

\begin{figure}[!t]
\centering
\centerline{\includegraphics[width=\columnwidth]
{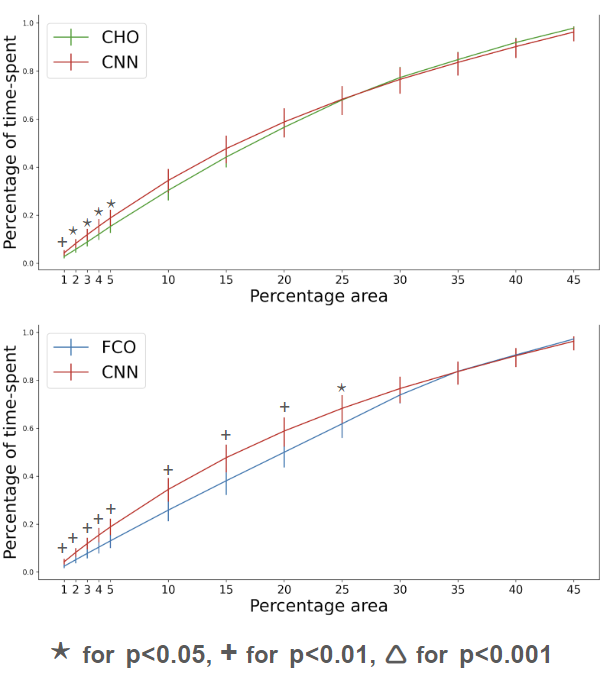}}
\caption{
\textbf{Mass:} Percentage of time spent corresponding to the top locations in the model observer response map. Time-Spent corresponding to top $1\%$, $5\%$, $10\%$, $15\%$, $20\%$, $25\%$, $30\%$, $35\%$, $40\%$, $45\%$,  locations of the model observer response map
}
\label{time_spent_analysis_mass}
\end{figure}

Figure~\ref{time_spent_analysis} and Figure~\ref{time_spent_analysis_mass} show the percentage of time radiologists spent fixating the regions with the highest model observer response scores. 
We repeat the analysis for both types of signals (microcalcification and mass), and the corresponding results for mass, which produced significant results, are shown in Figure~\ref{time_spent_analysis_mass}. 
Figure~\ref{time_spent_analysis} shows the time spent by the radiologists corresponding to the top $1\%$ locations of the model observer response map.
For the model observers trained for detecting microcalcification, radiologists fixated longer in the top locations corresponding to the CNN response map than CHO (p = $0.082$) and FCO ($p = 0.120$), but it is not significant. 
For the model observers trained for detecting mass, radiologists fixated longer at the top locations corresponding to the CNN, and the difference is significant ($p = 0.009$ for CHO and $p = 0.004$ for FCO). 
For the bottom row, the same analysis is repeated for many different percentages along with $1\%$. The percentage of time spent reaches $100\%$ as a large percentage of the area is considered.
Overall, top locations from CNN coincided with the location where more time was spent by radiologists showing potential similarity between CNN's false positive locations and false positives of the human template.


\begin{figure}[!t]
\centerline{\includegraphics[width=0.8\columnwidth]{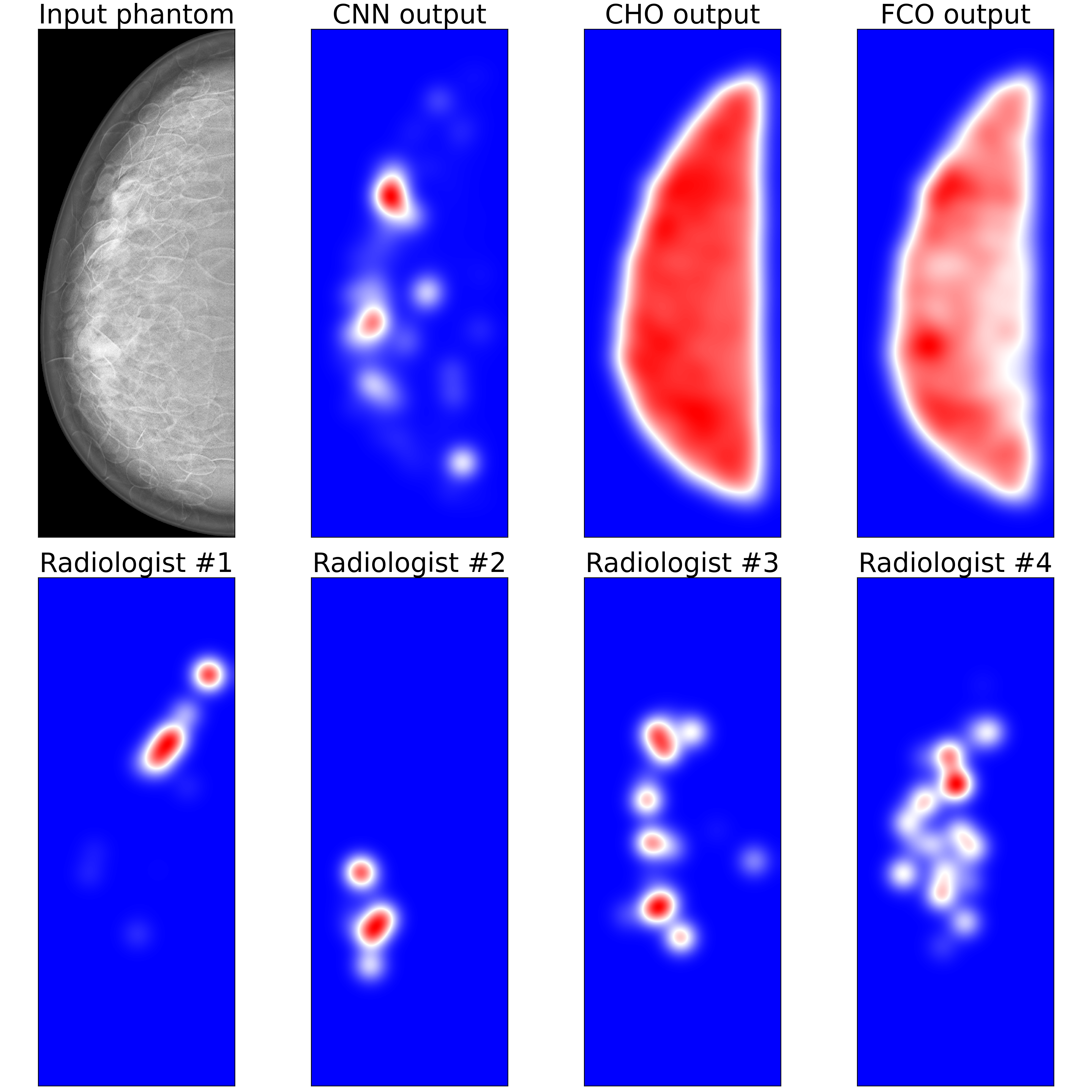}}
\caption{Visualization of model output maps and fixation maps of radiologists. \textbf{First row:} Input phantom slice and the corresponding output of the model observers. \textbf{Second row:} Fixation map of four radiologists on this phantom slice.}
\label{fig3}
\end{figure}


Therefore, CNN is a significantly better predictor of where humans will spend longer. It is an advantage that can be explored in the future as it adds a new dimension to the usual performance benefit seen in CNNs.

\section{Discussion}


\subsection{Limitations of linear model observers for search tasks:}
Our work extends the evaluation of linear model observers from simpler LKE tasks common in the medical imaging field to more clinically realistic search tasks. 

Our main finding shows that linear model observers that have worked well to evaluate and optimize task-based medical image quality for LKE tasks~\cite{p5, p7, p8, Castella2009MassDO, Eckstein2000APG, Daz2015DerivationOA} fail to achieve an accuracy comparable to human performance (radiologists) in a subset of  more clinically realistic search tasks (2D microcalcification and 3D mass search). This finding might justify the recent adoption of CNN model observers to optimize image quality ranging from ultrasound~\cite{Zhang2021CNNBasedMU}, MR~\cite{Fantini2021AutomaticMI} to teleophthalmology~\cite{Wang2019ACR}.

\subsection{CNNs learn to discount anatomical structures, unlike linear model observers:}

Why do the linear model observers underperform in the search task? We hypothesize that linear model observer template responses generate many false positive locations in the search. The linear models treat the background as a Gaussian process and thus cannot learn to dampen responses to the higher-order moments in the DBT phantom~\cite{Eckstein2006TheEO}. 

On the second front, we investigate the predictive power of a CNN-based model observer in predicting the amount of time spent by radiologists.
False positive locations for humans are the locations that look similar to the signal.
Since these false positive locations look similar to the signal, they tend to spend longer at these locations before moving to a new location.
From the model observer's perspective, if a signal-absent location looks similar to the signal, it produces a stronger response.
Therefore, if a location looks similar to the target, humans fixate longer at those locations, and model observers produce a stronger response.
This is the rationale behind comparing the time spent on maps of radiologists against the response maps of the model observers.
This analysis was done 
by picking the top response locations and seeing how much time was spent at those locations. By picking the top $1\%$ locations, we showed that radiologists spent significantly longer at the top locations from CNN.
We repeated this process both with the microcalcification and mass response maps. The significant results are seen only with the mass response maps implying that locations looking like mass-signal are more confusing for radiologists, who tend to spend more time at these locations.

\subsection{Limitations of CNNs as model observers for 3D search:}

Our results also show two discrepancies between the CNN model and the radiologists' performance. First, the accuracy of detecting the microcalcification is higher than the masses in the 3D search. In contrast, radiologists' accuracy is lower for the microcalcification than the masses in 3D search.  The second discrepancy is that the CNN model does not capture the accuracy degradation in detecting the microcalcification in 3D relative to 2D search.  Previous studies have shown that trained non-radiologists and radiologists' low accuracy for detecting microcalcifications in 3D search is related to search errors (not fixating the signal)~\cite{Lago2021UnderexplorationOT}. These search errors arise because of the low detectability of small signals when they are processed by points away from fixation (visual periphery) and observers' tendency to under-explore the 3D image stack with eye movements. Recent model observers incorporating visual processing foveal (at fixation) and periphery can predict these effects on human search accuracy~\cite{Lago2020FoveatedMO, Lago2021MedicalIQ}.

\section{Conclusion}
With the advancements in simulation techniques, we can now generate realistic phantoms at a lesser cost~\cite{Badano2018EvaluationOD, Bakic2014RealisticSO, Norris2014ASO}.
Convolution neural networks are now frequently applied to phantoms for task-based evaluation and optimization of medical image quality.  Our study demonstrates when and why CNNs outperform traditional linear model observers.  For location known exactly tasks, CNNs and linear observers attain similar accuracy but for search tasks linear observers' accuracy can fall below that of  CNNs and radiologists.  The linear observers' lower search accuracy is related to the lower ability, relative to CNNs and radiologists, to discount background false positives.

\end{document}